\documentclass[11pt]{article}

\usepackage[T1]{fontenc}
\usepackage[utf8]{inputenc}
\usepackage{lmodern}
\usepackage[margin=1in]{geometry}
\usepackage{amsmath,amssymb,amsthm,mathtools}
\usepackage{dsfont}
\usepackage{graphicx}
\usepackage{float}
\usepackage{tikz}
\usepackage{enumitem}
\usepackage[authoryear,round]{natbib}
\usepackage[colorlinks=true,linkcolor=blue!60!black,citecolor=blue!60!black,urlcolor=blue!60!black]{hyperref}
\usepackage[capitalise,noabbrev]{cleveref}

\theoremstyle{plain}
\newtheorem{theorem}{Theorem}
\newtheorem{lemma}[theorem]{Lemma}

\newtheorem{corollary}[theorem]{Corollary}
\theoremstyle{definition}
\newtheorem{definition}[theorem]{Definition}
\newtheorem{observation}[theorem]{Observation}
\crefname{observation}{Observation}{Observations}
\theoremstyle{remark}

\AddToHook{env/lemma/begin}{\crefalias{theorem}{lemma}}
\AddToHook{env/proposition/begin}{\crefalias{theorem}{proposition}}
\AddToHook{env/corollary/begin}{\crefalias{theorem}{corollary}}
\AddToHook{env/definition/begin}{\crefalias{theorem}{definition}}
\AddToHook{env/observation/begin}{\crefalias{theorem}{observation}}
\AddToHook{env/remark/begin}{\crefalias{theorem}{remark}}

\newcommand{\RANKING}{\texttt{RANKING}}
\newcommand{\RANDOM}{\texttt{RANDOM}}
\newcommand{\CUT}{\mathrm{CUT}}
\newcommand{\OPT}{\mathrm{OPT}}
\newcommand{\ALG}{\mathrm{ALG}}
\newcommand{\E}{\mathbb{E}}
\renewcommand{\P}{\mathbb{P}}
\newcommand{\1}{\mathds{1}}
\newcommand{\cF}{\mathcal{F}}

\newcommand{\tstar}{t^{*}}
\newcommand{\convP}{\xrightarrow{\mathbb{P}}}

\title{A Short Proof of the \texorpdfstring{$1-1/e$}{1-1/e} Performance of \texttt{RANKING}\\ on the Complete Upper Triangular Instance}

\author{
  Giovanni Montanari\thanks{FairPlay joint team (Inria, ENSAE, Criteo).}
  \and Nadav Merlis\thanks{Technion -- Israel Institute of Technology.}
  \and Simon Mauras\thanks{FairPlay joint team (Inria, ENSAE, Criteo); Inria.}
  \and Vianney Perchet\thanks{FairPlay joint team (Inria, ENSAE, Criteo); ENSAE Paris; Criteo AI Lab.}
}

\date{}

\begin{document}

\maketitle

\begin{abstract}
The randomized online bipartite matching algorithm \RANKING{} of Karp, Vazirani and Vazirani (1990) matches, asymptotically, a fraction $1-1/e$ of the vertices of the complete upper triangular (CUT) instance. We give a short proof of this fact that looks at the final matching from the offline side. A single observation drives the argument: the first $k$ arrivals can distinguish the offline vertices $v_k,\ldots,v_n$ only through their ranks. As a consequence, whether $v_k$ is matched is decided by a uniform draw among the remaining ranks, and the number of matched vertices among $v_1,\ldots,v_k$ is a Markov chain with explicit transition probabilities. A linear differential equation, solved by hand, gives the limiting profile of this chain, and the constant $1/e$ appears where the matching probability reaches one. The main text is elementary; all routine verifications are deferred to the appendix.

\end{abstract}

\section{Introduction}
\label{sec:intro}

In the online bipartite matching problem, we are given a bipartite graph $G=(U,V,E)$, where $V$ is a set of \emph{offline} vertices known in advance and $U$ is a set of \emph{online} vertices revealed one at a time. When an online vertex $u\in U$ arrives, its incident edges are revealed, and the algorithm must immediately and irrevocably either match $u$ to an available neighbor in $V$ or leave it unmatched. The goal is to maximize the size of the final matching. The problem models allocation tasks in which demand arrives sequentially and resources must be committed on the spot, such as online advertising \citep{MEHTA,GOEL}, kidney exchange \citep{KIDNEY,KIDNEY2} and student--school assignment \citep{AZEVEDO,LESHNO}.

The standard performance measure for online matching algorithms is the competitive ratio.

\begin{definition}[Competitive ratio]
An \emph{instance} $I$ consists of a bipartite graph together with an arrival order of the online vertices. Let $\ALG(I)$ be the size of the matching produced by an online algorithm on $I$, and $\OPT(I)$ the size of a maximum matching of the graph. The competitive ratio of the algorithm is
\[
\inf_I \frac{\E[\ALG(I)]}{\OPT(I)},
\]
where the expectation is taken over the internal randomness of the algorithm. The algorithm is \emph{$\alpha$-competitive} if $\E[\ALG(I)]\geq\alpha\,\OPT(I)$ for every instance $I$.
\end{definition}

In their seminal work, \citet{KARP} introduced the \RANKING{} algorithm and showed that it is $(1-1/e)$-competitive under adversarial arrivals, and that no randomized online algorithm achieves a better competitive ratio. The analysis of \RANKING{} has since been simplified and revisited several times \citep{GOEL,BIRNBAUM,DEVANUR_RANKING}. When online vertices are drawn from a known distribution, the ratio $1-1/e$ can be exceeded \citep{FELDMAN}.

\paragraph{The CUT instance.}
For balanced instances with a perfect matching, \citet{KARP} showed that the worst-case analysis of \RANKING{} can be restricted to upper triangular instances. Their impossibility result relies on a distribution over relabelings of the \emph{complete upper triangular} instance $\CUT_n$ and on Yao's principle. The analysis of this distribution goes through \RANDOM, which matches each arrival to an available neighbor chosen uniformly at random, and through Kurtz's theorem on the deterministic approximation of Markov jump processes \citep{KURTZ}. They also observe that \RANDOM{} and \RANKING{} have the same expected matching size on $\CUT_n$, so the asymptotic matching size of \RANKING{} on $\CUT_n$ is $(1-1/e)n+o(n)$.

\paragraph{This paper.}
We give a short, self-contained proof that \RANKING{} itself matches $(1-1/e)n+o(n)$ vertices of $\CUT_n$ in expectation. Instead of following the online arrivals, we look at the final matching from the offline side, and count the number $N(k)$ of matched vertices among $v_1,\ldots,v_k$. The argument has three steps.
\begin{enumerate}[leftmargin=1.8em]
\item \emph{One observation} (\cref{obs:rank-order}). The first $k$ arrivals are all adjacent to $v_k,\ldots,v_n$, so \RANKING{} can tell these vertices apart only through their ranks.
\item \emph{A Markov chain} (\cref{lemma:transition}). As a consequence, whether $v_k$ is matched is decided by a uniform draw among the remaining ranks, and $N(k)$ is a Markov chain with transition probability $\min\{1,(k-N(k-1))/(n-k+1)\}$.
\item \emph{A differential equation} (\cref{sec:main}). For large $n$, this chain follows a deterministic curve, given by a linear differential equation that we solve by hand. The constant $1/e$ appears where the matching probability reaches one.
\end{enumerate}
The rigorous counterparts of the third step, \cref{thm:cut-limit} and \cref{coroll:cut-performance}, follow from the differential equation method of \citet{WORMALD}. Besides the final value, the argument gives the whole profile of $N$: about $n/e$ offline vertices remain unmatched, and all but $o(n)$ of them lie among the first $(1-1/e)n$.

We kept the main text elementary on purpose: each step can be checked by hand, and the whole argument can be read in a few minutes. The remaining verifications are routine and are deferred to the appendix, together with a proof of the version of the differential equation method that we use and numerical simulations. \Cref{sec:model} recalls \RANKING{} and $\CUT_n$, \cref{sec:exploration,sec:main} contain the three steps, and \cref{sec:conclusion} concludes.

\section{The algorithm and the instance}
\label{sec:model}

\RANKING{} draws a uniformly random ordering of the offline vertices once, before any arrival, and keeps it fixed for the whole run.

\begin{figure}[H]
\fbox{\begin{minipage}{\dimexpr\linewidth-2\fboxsep-2\fboxrule\relax}
\textbf{Algorithm} \RANKING
\begin{itemize}[leftmargin=1.5em]
\item \textbf{Initially known:} the offline vertex set $V$, with $n=|V|$.
\item \textbf{Initialization:} assign the offline vertices distinct ranks from $1$ to $n$, uniformly at random among the $n!$ possible assignments. Denote the rank of $v_j$ by $r_j$; smaller ranks indicate higher priority.
\item \textbf{Upon arrival of $u\in U$:} if $u$ has an unmatched neighbor, match $u$ to the unmatched neighbor of smallest rank; otherwise leave $u$ unmatched.
\end{itemize}
\end{minipage}}
\end{figure}

Using the same random ranking throughout the run gives \RANKING{} a better worst-case guarantee than any deterministic online algorithm and than \RANDOM{} \citep{KARP}.

\begin{figure}[htbp]
\centering
\includegraphics{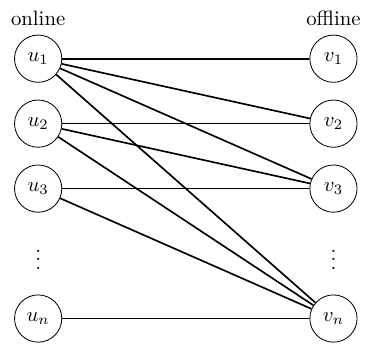}
\caption{The complete upper triangular instance $\CUT_n$. Online vertex $u_i$ is adjacent to $v_i,\ldots,v_n$.}
\label{fig:cut}
\end{figure}

Throughout, $U=\{u_1,\ldots,u_n\}$ and $V=\{v_1,\ldots,v_n\}$, and the online vertices arrive in the order $u_1,\ldots,u_n$. An \emph{upper triangular} instance contains all diagonal edges $(u_i,v_i)$, may contain edges $(u_i,v_j)$ with $i<j$, and contains no edge with $i>j$: its adjacency matrix, with rows indexing online vertices and columns indexing offline vertices, is upper triangular. The \emph{complete upper triangular instance} $\CUT_n$ has edge set
\[
E=\{(u_i,v_j):1\leq i\leq j\leq n\}.
\]
The neighborhood of $u_i$ is $\{v_i,\ldots,v_n\}$, so online neighborhoods shrink as the index increases. Conversely, $v_j$ is adjacent exactly to the first $j$ online vertices $u_1,\ldots,u_j$. The diagonal edges form a perfect matching, so $\OPT(\CUT_n)=n$; see \cref{fig:cut}.

Early online vertices have many available neighbors, but choosing an offline vertex with a large index may consume a resource needed by a later arrival. Although the graph admits a perfect matching, an online algorithm does not know which offline vertices will be needed by future arrivals and may therefore leave later vertices unmatched. This imbalance between the flexibility of early arrivals and the scarcity faced by later ones is what makes $\CUT_n$ hard.

\section{One observation, and a Markov chain}
\label{sec:exploration}

Fix the random ranking $r=(r_1,\ldots,r_n)$ and let $M_r$ be the final matching returned by \RANKING{} on $\CUT_n$. We look at this matching from the offline side, one offline vertex at a time:
\[
X_k=\1\{v_k\text{ is matched in }M_r\},
\qquad N(0)=0,\qquad N(k)=\sum_{j=1}^k X_j .
\]
Thus $N(k)$ counts the matched vertices among $v_1,\ldots,v_k$, and $N(n)=|M_r|$. The index $k$ refers to offline vertices, not to the number of online arrivals processed so far. The whole analysis rests on the following observation.

\begin{observation}
\label{obs:rank-order}
Let $S_k=\{v_k,\ldots,v_n\}$. Every vertex of $S_k$ is adjacent to every one of the first $k$ arrivals $u_1,\ldots,u_k$. During these arrivals, \RANKING{} can therefore distinguish the vertices of $S_k$ only through their ranks: the vertices of $S_k$ that it matches are those of smallest rank in $S_k$.
\end{observation}

\begin{figure}[htbp]
\centering
\includegraphics{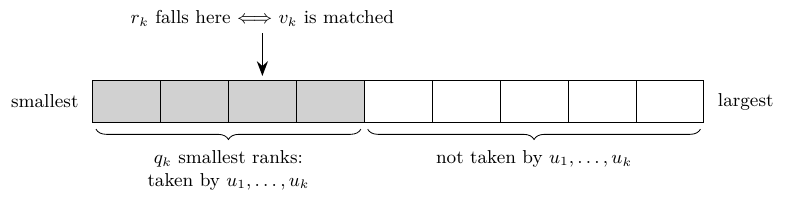}
\caption{The criterion~\eqref{eq:cut-rank-criterion}. The ranks still carried by $S_k=\{v_k,\ldots,v_n\}$ are sorted in increasing order. The first $k$ arrivals take the vertices holding the $q_k$ smallest of them, and $v_k$ is matched exactly when its rank $r_k$ falls in this block.}
\label{fig:rank-criterion}
\end{figure}

Since $v_k$ has no neighbor after $u_k$, it is matched if and only if it is one of these vertices. It remains to count them. Let $q_k$ be the number of vertices of $S_k$ matched during the first $k$ arrivals. A vertex $v_j$ with $j<k$ can only be matched to one of $u_1,\ldots,u_j$, so exactly $N(k-1)$ of the first $k$ arrivals are matched outside $S_k$. If some vertex of $S_k$ is still free after $u_k$, it was available to each of $u_1,\ldots,u_k$, so all of them were matched and $q_k=k-N(k-1)$. Otherwise $q_k=n-k+1$. In both cases $q_k=\min\{k-N(k-1),\,n-k+1\}$, and
\begin{equation}\label{eq:cut-rank-criterion}
X_k=1
\quad\Longleftrightarrow\quad
r_k\text{ is among the }q_k\text{ smallest ranks in }S_k.
\end{equation}
\Cref{fig:rank-criterion} illustrates this criterion.

Now reveal the ranks $r_1,r_2,\ldots$ one at a time, with $\cF_k=\sigma(r_1,\ldots,r_k)$. After $r_1,\ldots,r_{k-1}$ are revealed, we know $N(k-1)$, and we know which ranks remain for $S_k$, but not which vertex holds which. The rank $r_k$ is then a uniform draw among these $n-k+1$ ranks, as from an urn of $n-k+1$ balls of which $q_k$ are winning.

\begin{lemma}[Conditional matching probability]
\label{lemma:transition}
For $1\leq k\leq n$, the variable $N(k)$ is $\cF_k$-measurable and
\begin{equation}\label{eq:cut-transition}
\P(X_k=1\mid\cF_{k-1})
=\min\left\{1,\frac{k-N(k-1)}{n-k+1}\right\}.
\end{equation}
\end{lemma}

\begin{proof}
Measurability follows from~\eqref{eq:cut-rank-criterion} by induction on $k$ (\cref{app:lemma}). Given $\cF_{k-1}$, the set $R_k=\{1,\ldots,n\}\setminus\{r_1,\ldots,r_{k-1}\}$ of ranks carried by $S_k$ and the number $q_k$ are known, and $r_k$ is uniform on $R_k$ because the ranking is uniform. By~\eqref{eq:cut-rank-criterion}, exactly $q_k$ of the $n-k+1$ possible values give $X_k=1$, so $\P(X_k=1\mid\cF_{k-1})=q_k/(n-k+1)$.
\end{proof}

For example, $v_1$ is matched exactly when $r_1=1$, which has probability $1/n$. For $n\geq2$, conditionally on $N(1)=1$, the vertex $v_2$ must carry the smallest remaining rank, which has probability $1/(n-1)$. Conditionally on $N(1)=0$, it must carry one of the two smallest remaining ranks, which has probability $\min\{1,2/(n-1)\}$. If $r_1=2$, for instance, the two smallest remaining ranks are $1$ and $3$.

The conditional law of $X_k$ depends only on $k$ and $N(k-1)$, so $(N(k))_{0\leq k\leq n}$ is a time-inhomogeneous Markov chain. Its matching probability is nondecreasing in $k$ along every realization (\cref{app:lemma}): once it reaches one, every remaining offline vertex is matched.

\section{From the Markov chain to a differential equation}
\label{sec:main}

Since $\OPT(\CUT_n)=n$, the quantity $\E[N(n)]/n$ is the expected performance ratio of \RANKING{} on $\CUT_n$. \Cref{lemma:transition} describes the exploration one step at a time; for large $n$ these steps add up to a deterministic curve, which we first derive informally.

\paragraph{Heuristic derivation.}
Write $t=k/n$ and suppose that $N(k)\approx n\,z(t)$ for some smooth function $z$. One step of the exploration moves $t$ by $1/n$ and, by \cref{lemma:transition}, increases $N$ in expectation by
\[
\E[N(k+1)-N(k)\mid\cF_k]
=\min\left\{1,\frac{k+1-N(k)}{n-k}\right\}
\approx\min\left\{1,\frac{t-z(t)}{1-t}\right\}.
\]
On the other hand, $N(k+1)-N(k)\approx n\bigl(z(t+1/n)-z(t)\bigr)\approx z'(t)$. Equating the two gives
\begin{equation}\label{eq:ode-main}
z'(t)=\min\left\{1,\frac{t-z(t)}{1-t}\right\},
\qquad z(0)=0.
\end{equation}
The right-hand side is the probability that the offline vertex at position $t$ is matched. As long as it is below one, the equation is linear and can be written as
\[
\left(\frac{z(t)}{1-t}\right)'=\frac{t}{(1-t)^2},
\qquad\text{hence}\qquad
z(t)=t+(1-t)\log(1-t).
\]
Along this solution, $(t-z(t))/(1-t)=-\log(1-t)$. The matching probability therefore reaches one exactly when $1-t=1/e$, that is, at $\tstar=1-1/e$. From then on every offline vertex is matched, $z$ grows with slope one, and $z(t)=z(\tstar)+(t-\tstar)=t-1/e$. At $t=1$ this gives the matching size $1-1/e$.

\begin{figure}[htbp]
\centering
\includegraphics[width=0.85\linewidth]{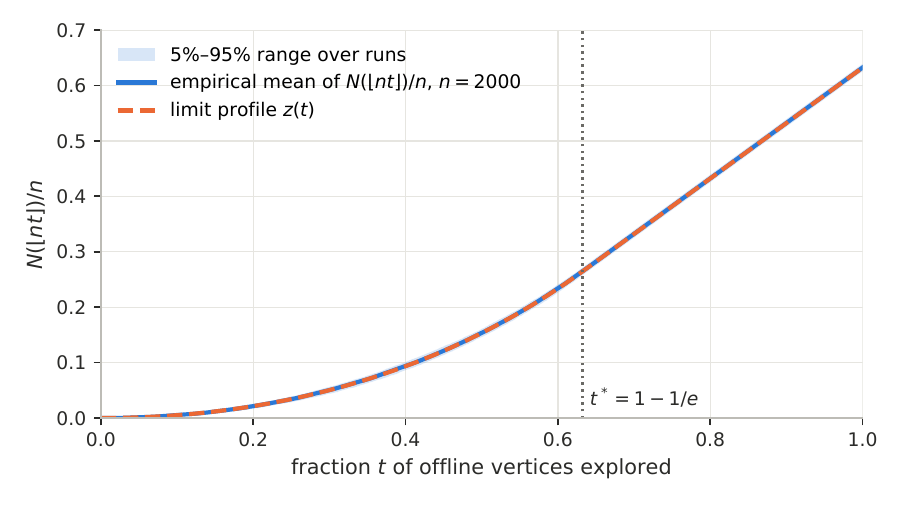}
\caption{Exploration profile $N(\lfloor nt\rfloor)/n$ for $n=2000$: empirical mean and $5\%$--$95\%$ range over $200$ runs, compared with the limit profile $z(t)$ of \cref{thm:cut-limit}. The dotted line marks $\tstar=1-1/e$.}
\label{fig:profile}
\end{figure}

\Cref{fig:profile} compares this curve with simulated exploration profiles. About $n/e$ offline vertices remain unmatched, and all but $o(n)$ of them lie among the first $\tstar n$: almost every later vertex is matched.

\paragraph{Rigorous statements.}
This heuristic can be made rigorous with the differential equation method of \citet{WORMALD}. The increments of $N$ are bounded by one, and on $[0,T]$ with $T<1$ the expected increment differs from a Lipschitz function of $(k/n,N(k)/n)$ by $O(1/n)$. These are the conditions of \cref{thm:wormald}, which we verify in \cref{app:proof-limit}.

\begin{theorem}[Limiting profile]
\label{thm:cut-limit}
Let $N(k)$ be the number of matched vertices found after $k$ stages of the offline exploration of \RANKING{} on $\CUT_n$, and let $\tstar=1-1/e$. For every fixed $T\in(0,1)$, as $n\to\infty$,
\[
\sup_{0\leq t\leq T}\left|\frac{N(\lfloor nt\rfloor)}{n}-z(t)\right|\convP 0,
\]
where
\begin{equation}\label{eq:cut-profile}
z(t)=
\begin{cases}
t+(1-t)\log(1-t), & 0\leq t\leq \tstar,\\
t-1/e, & \tstar\leq t<1.
\end{cases}
\end{equation}
\end{theorem}

The theorem stops short of $t=1$, where the drift is not Lipschitz. The endpoint follows because the last $n-\lfloor nT\rfloor$ steps change $N$ by at most $n(1-T)+1$, and $T$ can be taken arbitrarily close to one (\cref{app:proof-corollary}).

\begin{corollary}[Performance ratio]
\label{coroll:cut-performance}
Let $N(n)$ be the size of the matching returned by \RANKING{} on $\CUT_n$. As $n\to\infty$,
\[
\frac{N(n)}{n}\convP 1-\frac1e,
\qquad
\lim_{n\to\infty}\frac{\E[N(n)]}{n}=1-\frac1e.
\]
\end{corollary}

By \cref{coroll:cut-performance}, the performance ratio on $\CUT_n$ converges to $1-1/e$, which is also the competitive ratio of \RANKING{} \citep{KARP}. The family $(\CUT_n)_{n\geq1}$ is therefore asymptotically worst-case for \RANKING.

\section{Conclusion}
\label{sec:conclusion}

The asymptotic performance of \RANKING{} on the complete upper triangular instance follows from one observation: the first $k$ arrivals see $v_k,\ldots,v_n$ only through their ranks. This turns the offline side of the matching into an urn, and the urn into a differential equation that can be solved by hand.

The argument suggests a general way to analyze an online matching algorithm on a given family of graphs: explore the offline vertices one at a time and follow the number of matched vertices found so far. All the work is in one quantity, the conditional probability that the next offline vertex is matched given the exploration so far. Once this probability is known as a function of a few aggregate statistics, the differential equation method gives the matching profile, and the profile gives the performance ratio. On $\CUT_n$ this probability has the closed form of \cref{lemma:transition}. On other instances it may only be bounded; bounds of the same form would then translate, through the same differential equation, into bounds on the performance ratio.

\section*{Acknowledgements}

\begin{sloppypar}
This work was supported in part by the French National Research Agency (ANR) through grants ANR-20-CE23-0007 and ANR-23-CE23-0002, and through the PEPR IA FOUNDRY project (ANR23-PEIA-0003). It was also supported by Hi! PARIS and the ANR/France 2030 program (ANR23-IACL-0005).
\end{sloppypar}

\bibliographystyle{plainnat}
\bibliography{references}

\appendix
\section{Details for \texorpdfstring{\cref{sec:exploration}}{Section 3}}
\label{app:lemma}

\paragraph{Measurability in \texorpdfstring{\cref{lemma:transition}}{Lemma 3}.}
We show by induction on $k$ that $N(k)$ is $\cF_k$-measurable, with $\cF_0=\{\varnothing,\Omega\}$. The initial value $N(0)=0$ is deterministic. Suppose that $N(k-1)$ is $\cF_{k-1}$-measurable. The ranks $r_1,\ldots,r_{k-1}$ determine the set
\[
R_k=\{1,\ldots,n\}\setminus\{r_1,\ldots,r_{k-1}\}
\]
of ranks carried by the vertices of $S_k$; only their assignment to individual vertices is unknown. Hence $q_k=\min\{k-N(k-1),n-k+1\}$ and the $q_k$ smallest elements of $R_k$ are $\cF_{k-1}$-measurable. Once $r_k$ is revealed, criterion~\eqref{eq:cut-rank-criterion} determines $X_k$, so $X_k$ and $N(k)=N(k-1)+X_k$ are $\cF_k$-measurable.

\paragraph{Monotonicity of the matching probabilities.}
Write
\[
p_k=\P(X_k=1\mid\cF_{k-1})=\min\left\{1,\frac{k-N(k-1)}{n-k+1}\right\}.
\]
Since $N(k)=N(k-1)+X_k$ with $X_k\in\{0,1\}$,
\[
k+1-N(k)=k-N(k-1)+1-X_k\geq k-N(k-1).
\]
From $p_k$ to $p_{k+1}$, the numerator therefore stays the same (if $X_k=1$) or increases by one (if $X_k=0$), while the denominator decreases from $n-k+1$ to $n-k$, both positive for $1\leq k<n$. As the numerators are nonnegative,
\[
\frac{k+1-N(k)}{n-k}\geq\frac{k-N(k-1)}{n-k}\geq\frac{k-N(k-1)}{n-k+1},
\]
and taking the minimum with $1$ gives $p_{k+1}\geq p_k$. Once $p_k$ reaches $1$, it stays equal to $1$, so all remaining offline vertices are matched almost surely. The index at which this saturation occurs is random; \cref{thm:cut-limit} shows that, after rescaling, it concentrates around $\tstar=1-1/e$.

\section{A bounded-increment differential equation approximation}
\label{app:wormald}
We record the scalar, bounded-increment form of the differential equation method needed above. It is a specialized version of the method of \citet{WORMALD}; see also \citet{KURTZ}. We include a proof to keep the paper self-contained.

\begin{theorem}\label{thm:wormald}
Fix $T>0$. For each $n$, let $(Y_n(k))_{0\leq k\leq\lfloor nT\rfloor}$ be adapted to a filtration $(\mathcal H_{n,k})$. Suppose that, for constants $a<b$ and $B<\infty$ independent of $n$:
\begin{enumerate}
\item $Y_n(k)/n\in[a,b]$ almost surely for all relevant $k$, and $Y_n(0)/n\xrightarrow{\mathbb P}z_0$;
\item $|Y_n(k+1)-Y_n(k)|\leq B$ almost surely;
\item a Lipschitz function $f:[0,T]\times[a,b]\to\mathbb R$ and deterministic numbers $\delta_n\to0$ satisfy, almost surely,
\[
\left|\mathbb E[Y_n(k+1)-Y_n(k)\mid\mathcal H_{n,k}]
-f(k/n,Y_n(k)/n)\right|\leq\delta_n.
\]
\end{enumerate}
If the initial-value problem $z'(t)=f(t,z(t))$, $z(0)=z_0$, has a solution on $[0,T]$ taking values in $[a,b]$, then this solution is unique and
\[
\sup_{0\leq t\leq T}\left|\frac{Y_n(\lfloor nt\rfloor)}n-z(t)\right|
\xrightarrow{\mathbb P}0.
\]
\end{theorem}

\begin{proof}
Write $d_{n,j}=\mathbb E[Y_n(j+1)-Y_n(j)\mid\mathcal H_{n,j}]$ and define the martingale
\[
M_n(k)=\sum_{j=0}^{k-1}\bigl(Y_n(j+1)-Y_n(j)-d_{n,j}\bigr).
\]
Its increments have absolute value at most $2B$. Orthogonality of martingale increments and Doob's $L^2$ maximal inequality give
\[
\mathbb E\!\left[\max_{k\leq\lfloor nT\rfloor}|M_n(k)|^2\right]
\leq 16B^2\lfloor nT\rfloor.
\]
Consequently $\max_k|M_n(k)|/n\xrightarrow{\mathbb P}0$.

Set $y_{n,k}=Y_n(k)/n$. Summing the drift approximation yields
\[
y_{n,k}=y_{n,0}+\frac1n\sum_{j=0}^{k-1}f(j/n,y_{n,j})
+\frac{M_n(k)}n+R_{n,k},
\qquad |R_{n,k}|\leq T\delta_n.
\]
The function $f$ is bounded on its compact domain; hence $z$ is Lipschitz. The Lipschitz property of $f$ then bounds the error of the corresponding left Riemann sums uniformly in $k$:
\[
z(k/n)=z_0+\frac1n\sum_{j=0}^{k-1}f(j/n,z(j/n))+O(1/n).
\]
Let $L$ be a Lipschitz constant for $f$ in its second argument. Subtracting these identities gives
\[
|y_{n,k}-z(k/n)|\leq A_n+\frac Ln\sum_{j=0}^{k-1}|y_{n,j}-z(j/n)|,
\]
where
\[
A_n=|y_{n,0}-z_0|+\frac{\max_j|M_n(j)|}{n}+T\delta_n+O(1/n)
\xrightarrow{\mathbb P}0,
\]
with the $O(1/n)$ term chosen nonnegative. Discrete Gronwall gives
\[
\max_k|y_{n,k}-z(k/n)|\leq A_n \exp(LT).
\]
The difference between $z(t)$ and $z(\lfloor nt\rfloor/n)$ is uniformly $O(1/n)$, proving the result. Uniqueness follows from the same Lipschitz condition and the continuous Gronwall inequality.
\end{proof}

\section{Proof of \texorpdfstring{\cref{thm:cut-limit}}{Theorem 3}}
\label{app:proof-limit}

Fix $T\in(0,1)$. We apply the bounded-increment version of Wormald's differential equation method stated in \cref{thm:wormald} in \cref{app:wormald}. We verify its assumptions on $[0,T]$ and identify the limiting trajectory.

Introduce the normalized exploration times and process
\[
t_k=\frac{k}{n},
\qquad
Z_n(t)=\frac{N(\lfloor nt\rfloor)}{n},
\qquad 0\leq t\leq1.
\]
Here, $t$ represents the fraction of offline vertices explored, while $Z_n(t)$ counts the matched vertices in that prefix, divided by $n$. In particular,
\[
Z_n(t_k)=\frac{N(k)}n.
\]
Each exploration step has length $1/n$ in normalized time and changes $Z_n$ by either $0$ or $1/n$.

We now check the conditions of \cref{thm:wormald}.

\medskip
\noindent\textit{Initial condition and bounded state space.}
The process $N(k)$ is adapted to $(\mathcal F_k)$ by \cref{lemma:transition}. Moreover,
\[
Z_n(0)=\frac{N(0)}n=0,
\qquad
0\leq Z_n(t)\leq1.
\]
Thus the normalized initial condition is deterministic, and the normalized process takes values in the fixed interval $[0,1]$.

\medskip
\noindent\textit{Bounded increments.}
Since each stage of the exploration reveals the matched status of one offline vertex,
\[
N(k+1)-N(k)=X_{k+1}\in\{0,1\}.
\]
Consequently, the increments of the unnormalized process are bounded by $1$, independently of $n$.

\medskip
\noindent\textit{Approximation of the conditional drift.}
By \cref{lemma:transition},
\[
\mathbb E[N(k+1)-N(k)\mid\mathcal F_k]
=
\min\left\{1,\frac{k+1-N(k)}{n-k}\right\}.
\]
Using the normalized variables, we can rewrite the fraction as
\[
\frac{k+1-N(k)}{n-k}
=
\frac{t_k-Z_n(t_k)}{1-t_k}
+\frac{1}{n(1-t_k)}.
\]
Define
\[
f(t,x)=\min\left\{1,\frac{t-x}{1-t}\right\},
\qquad (t,x)\in[0,1)\times[0,1].
\]
We show that $f(t_k,Z_n(t_k))$ approximates the conditional drift uniformly up to time $T$.

To this end, let $h(a)=\min\{1,a\}$. For all real numbers $a,b$,
\[
|h(a)-h(b)|\leq|a-b|.
\]
Applying this inequality with
\[
a=\frac{t_k-Z_n(t_k)}{1-t_k}
+\frac{1}{n(1-t_k)},
\qquad
b=\frac{t_k-Z_n(t_k)}{1-t_k},
\]
gives
\[
\left|
\mathbb E[N(k+1)-N(k)\mid\mathcal F_k]
-f(t_k,Z_n(t_k))
\right|
\leq\frac{1}{n(1-t_k)}.
\]
For $0\leq k<\lfloor nT\rfloor$, we have $1-t_k\geq1-T>0$. Hence,
\begin{equation}\label{eq:cut-drift-error}
\left|
\mathbb E[N(k+1)-N(k)\mid\mathcal F_k]
-f\left(\frac{k}{n},\frac{N(k)}n\right)
\right|
\leq\frac{1}{n(1-T)}.
\end{equation}
Since $T$ is fixed, the right-hand side tends to zero uniformly over the relevant indices. This verifies the drift approximation condition with error
\[
\delta_n=\frac{1}{n(1-T)}.
\]

\medskip
\noindent\textit{Regularity of the limiting drift.}
We next verify that $f$ is Lipschitz on $[0,T]\times[0,1]$. Consider
\[
g(t,x)=\frac{t-x}{1-t}.
\]
On this rectangle, its partial derivatives satisfy
\[
\left|\frac{\partial g}{\partial x}(t,x)\right|
=\frac{1}{1-t}\leq\frac{1}{1-T},
\qquad
\left|\frac{\partial g}{\partial t}(t,x)\right|
=\frac{1-x}{(1-t)^2}\leq\frac{1}{(1-T)^2}.
\]
In particular, for $(t,x),(s,y)\in[0,T]\times[0,1]$,
\[
|g(t,x)-g(s,y)|
\leq
\frac{|t-s|}{(1-T)^2}
+\frac{|x-y|}{1-T}.
\]
Since $f=h\circ g$ and $h$ is $1$-Lipschitz, the same bound holds for $|f(t,x)-f(s,y)|$. This establishes the required regularity.

\medskip
\noindent\textit{The associated differential equation.}
The limiting equation is determined by the conditional rate of change of the normalized process. Indeed,
\[
\frac{
\mathbb E[Z_n(t_{k+1})-Z_n(t_k)\mid\mathcal F_k]
}{
t_{k+1}-t_k
}
=
\mathbb E[N(k+1)-N(k)\mid\mathcal F_k],
\]
because both the process increment and the time increment are divided by $n$. By~\eqref{eq:cut-drift-error}, this rate is approximated by $f(t_k,Z_n(t_k))$. The corresponding deterministic trajectory therefore satisfies
\begin{equation}\label{eq:cut-ode}
z'(t)=f(t,z(t))
=\min\left\{1,\frac{t-z(t)}{1-t}\right\},
\qquad z(0)=0.
\end{equation}
To complete the verification of the theorem's assumptions, we construct a solution on $[0,T]$ and check that it remains in $[0,1]$.

At the initial condition, $(t-z(t))/(1-t)=0$. We first solve the unsaturated branch of~\eqref{eq:cut-ode},
\[
z'(t)+\frac{z(t)}{1-t}=\frac{t}{1-t}.
\]
Multiplying by the integrating factor $(1-t)^{-1}$ gives
\[
\frac{d}{dt}\left(\frac{z(t)}{1-t}\right)
=\frac{t}{(1-t)^2}.
\]
Integrating from $0$ to $t$ and using $z(0)=0$, we obtain
\[
\frac{z(t)}{1-t}
=
\int_0^t\frac{s}{(1-s)^2}\,ds
=
\frac{1}{1-t}+\log(1-t)-1.
\]
Thus,
\[
z(t)=t+(1-t)\log(1-t).
\]

This expression solves the original equation as long as the argument of the minimum does not exceed $1$. Along this trajectory,
\[
\frac{t-z(t)}{1-t}=-\log(1-t),
\]
which increases from $0$ and first reaches $1$ at
\[
t^*=1-\frac1e.
\]
The unsaturated solution is therefore valid on $[0,t^*]$, and
\[
z(t^*)=1-\frac2e.
\]

For $t\geq t^*$, consider the continuation with derivative $1$:
\[
z(t)=z(t^*)+(t-t^*)=t-\frac1e.
\]
Along this continuation,
\[
\frac{t-z(t)}{1-t}
=\frac{1}{e(1-t)}\geq1,
\qquad t^*\leq t<1.
\]
Hence the minimum in~\eqref{eq:cut-ode} equals $1$, confirming that this continuation solves the equation. The two branches agree at $t^*$ and have the same derivative there. We have therefore obtained
\[
z(t)=
\begin{cases}
t+(1-t)\log(1-t),&0\leq t\leq t^*,\\
t-1/e,&t^*\leq t<1.
\end{cases}
\]
This is the profile stated in~\eqref{eq:cut-profile}. It is nondecreasing, starts at zero, and remains in $[0,1]$. The Lipschitz property of $f$ ensures uniqueness on every fixed interval $[0,T]$ with $T<1$.

All assumptions of \cref{thm:wormald} have now been verified. Consequently,
\[
\sup_{0\leq t\leq T}
\left|Z_n(t)-z(t)\right|
=
\sup_{0\leq t\leq T}
\left|\frac{N(\lfloor nt\rfloor)}n-z(t)\right|
\xrightarrow{\mathbb P}0.
\]

\section{Proof of \texorpdfstring{\cref{coroll:cut-performance}}{Corollary 4}}
\label{app:proof-corollary}
We first establish convergence of the normalized final matching size, and then pass to expectations.

\medskip
\noindent\textit{Convergence at the endpoint.}
Fix $T\in(t^*,1)$, where $t^*=1-1/e$, and set
\[
m_n=\lfloor nT\rfloor.
\]
By \cref{thm:cut-limit},
\[
\frac{N(m_n)}n\xrightarrow{\mathbb P}z(T)=T-\frac1e.
\]
After the first $m_n$ stages of the offline exploration, there are $n-m_n$ vertices left to examine. Each can contribute either zero or one additional matched vertex, so
\[
N(m_n)\leq N(n)\leq N(m_n)+n-m_n.
\]
Equivalently,
\begin{equation}\label{eq:cut-endpoint-bounds}
\underbrace{\frac{N(m_n)}n}_{L_n}
\leq
\frac{N(n)}n
\leq
\underbrace{\frac{N(m_n)}n+1-\frac{m_n}n}_{U_n}.
\end{equation}
By convergence at time $T$ and the fact that $m_n/n\to T$,
\[
L_n\xrightarrow{\mathbb P}T-\frac1e,
\qquad
U_n\xrightarrow{\mathbb P}
\left(T-\frac1e\right)+1-T
=1-\frac1e.
\]
For fixed $T$, the two limiting bounds differ by $1-T$. Since \cref{thm:cut-limit} applies to every fixed $T<1$, we can make this difference arbitrarily small by choosing $T$ sufficiently close to $1$.

To make this argument precise, put $c=1-1/e$ and fix $\varepsilon>0$. Choose
\[
T\in\left(
\max\left\{t^*,1-\frac{\varepsilon}{2}\right\},1
\right),
\]
and keep this $T$ fixed as $n\to\infty$. Define the event
\[
A_n=
\left\{
\left|\frac{N(m_n)}n-z(T)\right|
\leq\frac{\varepsilon}{4}
\right\}.
\]
\cref{thm:cut-limit} implies $\mathbb P(A_n)\to1$. Whenever $A_n$ occurs,
\[
z(T)-\frac{\varepsilon}{4}
\leq\frac{N(m_n)}n
\leq z(T)+\frac{\varepsilon}{4}.
\]
Substituting the lower inequality into~\eqref{eq:cut-endpoint-bounds} yields
\[
\frac{N(n)}n
\geq\frac{N(m_n)}n
\geq T-\frac1e-\frac{\varepsilon}{4}
=c-(1-T)-\frac{\varepsilon}{4}
>c-\varepsilon,
\]
where the last inequality uses $1-T<\varepsilon/2$.

Similarly, substituting the upper inequality gives
\[
\begin{aligned}
\frac{N(n)}n
&\leq\frac{N(m_n)}n+1-\frac{m_n}n\\
&\leq T-\frac1e+\frac{\varepsilon}{4}
      +1-\frac{m_n}n\\
&=c+\frac{\varepsilon}{4}
      +\left(T-\frac{m_n}n\right)\\
&<c+\frac{\varepsilon}{4}+\frac1n.
\end{aligned}
\]
Here we used
\[
0\leq T-\frac{\lfloor nT\rfloor}{n}<\frac1n.
\]
For all sufficiently large $n$, we have $1/n<3\varepsilon/4$, and hence
\[
\frac{N(n)}n<c+\varepsilon.
\]
Thus, for such $n$, the event $A_n$ implies
\[
\left|\frac{N(n)}n-c\right|<\varepsilon.
\]
Writing $A_n^c=\Omega\setminus A_n$ for the complementary event, we obtain
\[
\left\{
\left|\frac{N(n)}n-c\right|>\varepsilon
\right\}
\subseteq A_n^c.
\]
Therefore,
\[
\mathbb P\left(
\left|\frac{N(n)}n-c\right|>\varepsilon
\right)
\leq\mathbb P(A_n^c)
=1-\mathbb P(A_n)
\longrightarrow0.
\]
This proves
\[
\frac{N(n)}n\xrightarrow{\mathbb P}1-\frac1e.
\]

\medskip
\noindent\textit{Convergence of expectations.}
Set
\[
D_n=\left|\frac{N(n)}n-c\right|.
\]
Since both $N(n)/n$ and $c$ belong to $[0,1]$, we have $0\leq D_n\leq1$. For any $\eta>0$, split the expectation according to whether $D_n$ exceeds $\eta$:
\[
\begin{aligned}
\mathbb E[D_n]
&=\mathbb E[D_n\1_{\{D_n\leq\eta\}}]
 +\mathbb E[D_n\1_{\{D_n>\eta\}}]\\
&\leq\eta+\mathbb P(D_n>\eta).
\end{aligned}
\]
The first term is bounded by $\eta$ because $D_n\leq\eta$ on the corresponding event. The second is bounded by $\mathbb P(D_n>\eta)$ because $D_n\leq1$ everywhere.

Convergence in probability gives $\mathbb P(D_n>\eta)\to0$ for each fixed $\eta>0$. Consequently,
\[
0\leq\limsup_{n\to\infty}\mathbb E[D_n]\leq\eta.
\]
Since this holds for every $\eta>0$, letting $\eta\downarrow0$ yields
\[
\mathbb E[D_n]\longrightarrow0.
\]
This is precisely convergence of $N(n)/n$ to $c$ in $L^1$. Finally,
\[
\left|\frac{\mathbb E[N(n)]}{n}-c\right|
\leq
\mathbb E\left[\left|\frac{N(n)}n-c\right|\right]
\longrightarrow0,
\]
and therefore
\[
\lim_{n\to\infty}\frac{\mathbb E[N(n)]}{n}=1-\frac1e.
\]

Since $\mathrm{OPT}(\mathrm{CUT}_n)=n$, the quantity $\mathbb E[N(n)]/n$ is the expected performance ratio on $\mathrm{CUT}_n$.

\section{Numerical simulations}
\label{app:simulations}

We simulate \RANKING{} on $\CUT_n$ to illustrate the finite-size behavior of the quantities studied above; the convergence statements themselves are established by \cref{thm:cut-limit} and \cref{coroll:cut-performance}.

\begin{figure}[htbp]
\centering
\includegraphics[width=0.85\linewidth]{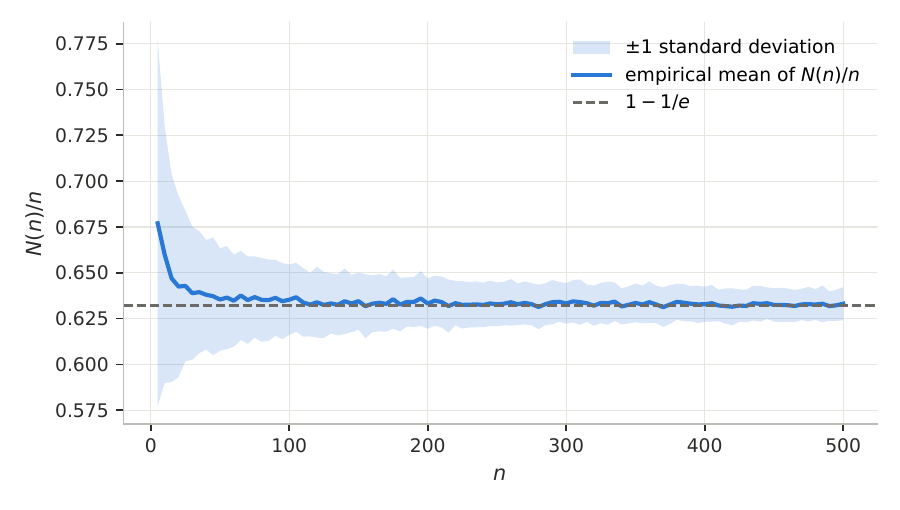}
\caption{Normalized matching size $N(n)/n$ of \RANKING{} on $\CUT_n$: empirical mean over $200$ runs for each $n$, with a band of $\pm1$ standard deviation across runs. The dashed line marks $1-1/e$.}
\label{fig:ratio}
\end{figure}

\Cref{fig:ratio} reports the empirical mean of $N(n)/n$ over $200$ independent runs for each $n\in\{5,10,\ldots,500\}$, together with a band of $\pm1$ standard deviation across runs. The mean approaches $1-1/e\approx0.632$ quickly, and the fluctuations shrink as $n$ grows, in agreement with the concentration of $N(n)/n$ in \cref{coroll:cut-performance}.

The exploration profile is shown in \cref{fig:profile} (\cref{sec:main}). For $n=2000$, the empirical mean over $200$ runs is indistinguishable from $z$ at that scale, and the $5\%$--$95\%$ range across runs is narrow.

\end{document}